
\magnification=1200

\hskip11cm IFT/9/1994
\vskip0.2cm

\centerline{\bf EQUATION OF STATE FOR PARTICLES ARISING AT THE UNIVERSE}
\centerline{\bf AT GRAND UNIFICATION ENERGIES}
\vskip0.3in
\centerline{ I. Dymnikova}
\vskip0.2in
\centerline{\sl N. Copernicus Astronomical Center}
\centerline{\sl Bartycka 18, PL-00-716 Warsaw, Poland}
\vskip0.1in
\centerline{\sl N. Copernicus Foundation for Polish Astronomy}
\centerline{\sl Al. Ujazdowskie 4, PL-00-478 Warsaw, Poland}
\vskip0.4in
\centerline{ M. Krawczyk}
\vskip0.2in
\centerline{\sl Institute of Theoretical Physics, Warsaw University}
\centerline{\sl Hoza 69, PL-00-681 Warsaw, Poland}
\vskip0.5in

\centerline{Abstract}

First postinflationary stage of the Universe evolution is considered
in more detail. It is shown that heavy particles with mass
$M_{H}\sim M_{GUT}$ arising at the Universe at  phase transitions
at Grand Unification Energies behave like ideal quantum degenerate Bose gas.
The equation of state for both scalar and gauge bosons is presented
including the coupling constant and vacuum expectation value at $E_{GUT}$.
One possible way is proposed to connect cosmological
observational data with parameters of Grand Unified Theories.
\vskip0.5in

The equation of state is necessary instrument for constructing the
cosmological models. Putting it into the Friedman equations, one gets
cosmological solution describing a certain stage of the evolution of the
Universe (see, e.g.${^{1}}$). In the case of an inhomogeneous situation,
the equations are more complicated, but procedure is on essence the same
(for review see ${^{2}}$).

On the other hand, the equation of state can also serve as useful
tool for testing physics of very high energies.
Calculating cosmological model with an equation of state at Grand
Unification energy $E_{GUT}\sim{10^{15}}$ GeV  one can hope to
get an opportunity
to extract some information concerning GUT physics from cosmological
observational data ${^{3}}$.

In the cosmological context the equation of state has the form
$$p = \alpha \varepsilon^{\gamma},\eqno(1)$$
where $p$ is pressure and $\varepsilon$ energy density of matter.
The case $\alpha=0$ corresponds to the dust equation of state,
the case $\gamma=1$ and $\alpha=1/3$ describes the ultrarelativistic
particles, and the case $\gamma=1$ and $\alpha=-1$ describes  vacuum
${^{4}}$
(or false vacuum ${^{5}})$.

The vacuum, or false vacuum, or inflationary equation of state
$$p=-\varepsilon\eqno(2)$$
has been proposed about thirty years ago.
In 1965, considering various possible equations
of state at superhigh densities, Sakharov ${^{6}}$ has proposed (2) as equation
of state for superdense matter. At the same time Gliner${^{4}}$  has proposed
(2) as the equation of state for vacuum with nonzero energy density,
because corresponding stress-energy tensor
$$T_{\alpha\beta}=\varepsilon g_{\alpha\beta}\eqno(3)$$
has an infinite set of comoving reference frames and no
distinguished one. Gliner has suggested that at superhigh densities a
superdense matter undergoes a transition into a state
of superdense vacuum. In 1970 he also suggested, that such a state could be
initial state for the expanding Universe with negative pressure as the
intrinsic reason for expansion ${^{7}}$.
Corresponding cosmological
model has been calculated in 1975 ${^{8}}$:
Because initial vacuum nonsingular state $(2)$ is homogeneous and isotropic,
and expansion is starting from the causally connected region confined by
the de Sitter event horizon $a_0$, here denoted by $r_0$:
$$r_{0}^{2}={{3c^4}\over{8\pi G\varepsilon}}\eqno(4)$$
the homogeneity and isotropy of the Universe have been guaranteed in
obvious way.

In 1981 Guth has found that eq.$(2)$ can arise in Grand Unified Theories
as the equation of state for scalar field in metastable state called false
vacuum ${^{5}}$, and now it is broadly used in inflationary scenarios
(for review see ${^{9,10}}$). Since 1981
many various and sophisticated physical
mechanisms have been proposed driving the inflation and
providing further Universe reheating (for review ${^{9,10}}$).
\vskip0.2in
In the present paper we consider the first postinflationary stage.
The state $(2)$ is unstable in both classical ${^{7}}$ and quantum ${^{11,12}}$
context. As a result of  phase transitions ${^{9}}$,
heavy nonrelativistic particles arise
with mass $M_{H}\sim M_{GUT}$ ${^{9,10}}$. Note, that
light particles can not live in the  causally connected region (4),
which at the energy scale $\sim 10^{15}$ GeV  is of a size $10^{-25}$ cm.

The first phase transition can be related to the spontaneous
global symmetry breaking
(because arrow of time is expected to appear just then),
hence at first scalar
bosons emerge at the Universe. So, we are starting with and focusing upon
scalar bosons. On the other hand,
all our estimates below are on essence valid
for both scalar and gauge bosons, and we present equation of state for both
of them.

First heavy particles emerging at the Universe can not be at rest,
at least because they are quantum particles, and they can undergo collisions.
As we do not expect that they appear with large momenta,
the CM energy of
their collisions $\sqrt s $
is close to the sum of rest energies of colliding particles
$\sqrt s \sim 2 E_H $ (the threshold of the reaction).
 Therefore the corresponding
interactions are mainly due to elastic 2 $\rightarrow$ 2 body scatterings.
 The production of one extra particle
will be possible at much higher CM energy, greater then
$\sqrt s_{inel} = 3 E_H$.

To describe our heavy particles by
the equation of state, we should consider them  as  gas.
To learn what kind of gas they form, we should determine characteristic
radius of interaction for them (depending on the coupling constant) and
compare it to geometrical radius, which for quantum particles is defined
by the Compton wavelength:
$$\lambda_{Compt}\sim {{\hbar}\over{M_{H}c}}\sim 10^{-29} cm.\eqno(5)$$
Corresponding effective geometrical cross-section is defined by
$$\sigma_{0} = \pi \lambda_{Compt}^2.\eqno(6)$$

Assuming that first heavy particles are scalars with the selfinteraction,
we can estimate their characteristic radius of interaction by the mass
of particles mediating the interaction. Since these particles are the
same scalars with masses $M_H$, the radiation of such a particle
for 2 $\rightarrow$ 2 body scatterings
is possible only due to the nonconservation of the energy
$$\Delta E \sim M_{H}c^2.\eqno(7)$$
Then, at the threshold, the cross section for the interaction mediated by
heavy particles and depending on  the coupling constant $g_H$
is given by

$$\sigma_{int}\sim {\alpha_H}^2/M_H^2\sim {\alpha_H}^2 \sigma_0,\eqno(8)$$
where $$ \alpha_H=g_H^2/4\pi.\eqno(9)$$

Now let us define the characteristic radius of interaction by the relation
$$\sigma_{int} = \pi r_{int}^2.\eqno(10)$$
Note , that $\sigma_{int}\ll\sigma_0$ , if $\alpha_H$ is small, thus
the characteristic radius of interaction in considered case is even
smaller than the Compton wavelength.

Heavy particles emerge in the vacuum with the energy density
$$\rho_{vac}=\rho_{GUT}=\rho_{Planck}({{M_{GUT}}\over{M_{Planck}}})^4
\sim 10^{77}g/cm^3.
\eqno(11)$$
The radius of causally connected region defined by the de Sitter formula $(4)$
is equal to $$r_{0}\sim10^{-25}cm.\eqno(12)$$
Hence we can estimate  the average momentum of the particles as

$$|\vec{p}| \sim \hbar /r_0 \sim 10^{12} GeV\eqno(13)$$

and average kinetic energy as
$$ E_{kin}=|\vec{p}|^2/2 M_H \sim \hbar^2/2 M_H r_0^2. $$
For the ratio we have

$$E_{kin}/{E_H} \sim \hbar^2/(r^2_0 M_H^2 c^2)\sim (\lambda_{Compt}/r_{0})^2
\sim 10^{-8}\eqno(14).$$
So, we can consider our heavy particles as dust, at the first
approximation. To tell more about them we should make sure that we have
ensemble of them to use statistics. This is just moment when we need some
cosmological model of transition, and now we know that we should organize
transition from vacuum (2) into the dust with pressure $p=0$
(up to $10^{-8}$ according to (14)).

\vskip0.2in

We describe the  phase transition using phenomenological cosmological
model. Following ${^{8,15}}$, let us assume that during a transition the
equation of state has the form
$$p + \varepsilon=\varepsilon_{1}{{(\varepsilon_{0}-\varepsilon)^{\beta}}
\over{(\varepsilon_{0}-\varepsilon_{1})^{\beta}}}\eqno(15)$$

At $\epsilon = \epsilon_{0}$ we have vacuum equation of state (2).
At $\epsilon = \epsilon_{1}$ we have dust equation of state $p$ = 0.
A parameter $0<\beta\leq 1$ characterizes the rate of transition, for
example, if a transition is driven by scalar field , $\beta$
characterizes the rate of rolling down a scalar field potentials with
small or nonexisting barrier between false and true vacuum.
For simplicity let us choose $\beta =1/2$ and
the case of the spatially-flat model ($\kappa = 0$).

Integrating the Friedmann equations with the transitional equation of state
$(15)$ we get for the scale factor $a$ the expression
$$a=a_{0}\exp\lbrace A\sin{{ct}\over{Aa_{0}}}\rbrace,\eqno(16)$$
where $$A=
2{{\sqrt{(1 -
\varepsilon_{1}/\varepsilon_{0}}}\over{3\varepsilon_{1}/\varepsilon_{0}}},\eqno(17)$$
and for the energy density the expression
$$\varepsilon=\varepsilon_{0}(1 - \sin^2{{ct}\over{Aa_{0}}}).\eqno(18)$$

Using condition of conservation of the rest mass at the end of the transition,
$a_{1}^3\varepsilon_{1}=a_{today}^3\varepsilon_{today}$, ${^{1}}$,
we get
$\varepsilon_{1}\sim(1/65)\varepsilon_{0}$,
$a_{1}\sim10^{-6}cm$ and duration of the transition
$t_{1}\sim 2\times10^{-34}sec$, adopting $\rho_{today}\sim10^{-29}g/cm^{3}$.

Because we organized a transition into the dust, the whole rest mass
($M_{1}\sim 2\times 10^{57}g$) is contained in our heavy particles with $M_H
\sim M_{GUT}\sim2\times10^{-9}g$. So, number of particles is
$N_{1}\sim 10^{66}$.
The volume at the end of transition is $V_{1}\sim 4\times 10^{-18}cm^3$,
hence
density of particles is

$$n={{N_{1}}\over{V_{1}}}\sim{10^{84}cm^{-3}}.\eqno(19)$$

Using the Friedmann equations and formulae (16)-(18), we can check the total
energy balance during the transition and find that only about $1/4$ of
vacuum energy involved in transition has been spent. The substantial amount
remains available for further expansion and reheating. Hence, the Universe
is still mainly vacuum dominated, and particles have at their disposal to
interact the causally connected region $(12)$. How many particles are in
this region?

$$N_{part. in \ caus.conn.reg.}=V_{caus.conn.reg}\times n$$

\centerline{$V_{caus.conn.reg.}=\pi^2r_o^3$    {\it (de Sitter volume).}}
So, $$N_{part. in\  caus.conn.reg.}\sim10^{10}.\eqno(20)$$
Hence, we have an ensemble and opportunity to use statistics.
\vskip0.2in

According to (19), the mean distance between particles is
$$r_{av}\sim 10^{-28}cm.\eqno(21)$$
The characteristic radius of interaction defined by (10),
$r_{int}\sim{\lambda_{Compt}\alpha_H}$ is much smaller than $r_{av}$.
Since the characteristic cross section
(8) is smaller than geometric cross-section (6), the free path
is determined by
$$l ={1\over{n\sigma_0}}\sim 3\times 10^{-27}cm.\eqno(22)$$
As a result our particles satisfy two general criteria for
ideal gas:

\vskip0.1in
$n r^{3}_{int}\ll 1~~~~~~\rightarrow$   First criterion for ideal gas\par

$r_{int}\ll l~~~~~~~~~ \rightarrow$  Second criterion for ideal gas\par
\vskip0.1in
Note, that the free path $l$ is much less than the radius of causally
connected region:
 $$l\ll r_0.\eqno(23)$$
Hence, in caussaly connected region particles can
collide and produce some temperature and pressure.

Next question is what kind of ideal gas we have.
The average momentum is given by (13), then the de Broglie wavelength
can be estimated as
$$\lambda_{de Broglie}={{\hbar}\over{\mid\vec{p}\mid}}\sim10^{-25}cm\eqno(23)$$
Because $\lambda_{de Broglie}\gg r_{av}$, our ideal gas is essentially quantum
one.
Indeed, the above condition  means that number of particles N is much more
than the number of quantum states at their disposal

$$\lbrace{{\lambda_{de Broglie}}\over{r_{av}}}\rbrace^3={{N}\over{V}}\cdot
{{\hbar^3}\over{p^3}}={{N}\over{(Vp^3)/\hbar^3}}\gg 1\eqno(24)$$
With $\mid\vec{p}\mid^2 \sim M_H k T_H$ where $k=1.38
\times10^{16}erg/K$ - the Boltzman constant, the condition (24) takes more
frequently used form ${^{16}}$

$${{N}\over{V}}{{\hbar^3}\over{(M_HkT_H)^{3/2}}}\gg1.\eqno(25)$$
So, we have ideal quantum Bose gas.

Its temperature of degeneration is determined by ${^{16}}$
$$k T_{deg}= 3.31 {{\hbar^2}\over{g_{st}^{2/3}M_H}} n^{2/3}.\eqno(26)$$
Here $g_{st}=2s+1$ is statistical weight, with $s=0$ for scalar and
$s=1$ for vector bosons.

For $M_H\sim2\times10^{-9}g$ and $n\sim 10^{84}cm^{-3}$
we obtain $k T_{deg}\sim10^{11}erg$.
For our gas with average momentum defined by (13),
$k T_H\sim E_{kin}\sim 3\times 10^4erg$.
Hence, its temperature $T_H\ll T_{deg}$,
and we can surely describe our gas by the equation of state for the
degenerate Bose gas ${^{16}}$
with the pressure given by
 $$P = 0.0851 g_{st}{{M_H^{3/2}}\over{\hbar^3}}
(kT_H)^{5/2}\eqno(27)$$

According to the standard approach (with selfinteraction $\lambda {\phi}^4$,
$\lambda=g_H^2$)
to generating masses of heavy scalars
via the spontaneous global symmetry breaking ${^{13,14}}$, we have

 $$M_H\sim  v_I g_H,\eqno(28)$$
where $v_I$ is a vacuum expectation value.
To reveal characteristic dependence of pressure on the coupling and the
vacuum expctation value, we can put i) average kinetic energy for $kT_{H}$ and
ii) express the quantity $r_{0}$ in terms of $M_{Pl}$ and $M_{GUT}$. Then
the equation (27) takes the form
$$P = const {{c^5}\over{\hbar^3}}{{1}\over{v_{I} g_{H}}} ({{M_{GUT}}\over
{M_{Pl}}})^5 M_{GUT}^{5}.\eqno(29)$$
We expect that $v_{I}\sim M_{GUT}$, so
$$P = const {{c^5}\over{\hbar^3}}{{1}\over{g_{H}}} ({{M_{GUT}}\over
{M_{Pl}}})^5 M_{GUT}^{4}.\eqno(30)$$

We see that in the equation of state for quantum ideal degenerate Bose gas
the coupling constant is appearing. Because, as it can be easily estimated,
pressure is much less than energy density,
the process of arising of particles
looks like evaporation of Bose-condensate.

It seems to be natural to expect that next in the earliest evolution
of the Univers  some gauge interaction
appears with  gauge bosons $X,Y$
and their gauge coupling constant
$g$. There are estimation
of the value of this gauge coupling at the energy scale $10^{15}$ GeV,
based on the accelerator's data at low energies ${^{17}}$. It gives for
the gauge coupling constant ${^{14,17}}$
$$\alpha=g^2/4\pi \sim 1/40-1/25.\eqno(31)$$

The spontaneous local symmetry breaking is expected to give a mass of these
gauge bosons of the order ${^{13,14}}$

$$M_X \sim {{g v_{II}} }.\eqno(31)$$

Since both effects are at similar scale $\sim {E}_{GUT}$,
and this is just scale of our analysis,
we can put $v_I \sim v_{II}$ and $g_H\sim g $.
(Note, that in SUSY case $v_{I} = v_{II} $ and $g_{H} =g$.)
 Hence the equation of state for the gauge bosons will be the same as for
scalar ones (see (27) or (29,30)) with appropriate change in constant.

The results presented here do not depend in a crucial way
whether the scale $M_{GUT}$ is equal to $10^15$ or
(what is preferable by SUSY models) $10^16$
GeV. There is also possible to have both phase transitions I and II
at the same time.

We can calculate cosmological model
including equation of state (29) depending on GUT parameters ${^{3}}$.
That way we could get opportunity to extract some information concerning GUT
physics from observational cosmological data, and compare it with
accelerator's data. It would provide some way toward astronomical testing
GUT models.
\vskip0.5in
\centerline{\bf Acknowledgments}
\vskip0.2in
One of us (I.D.) is fery grateful to R.de Ritis, C.Rubano and P.Scudellaro
for stimulating discussions concerning postinflationary stage.

This research was supported in part by the Polish Committee for Scientific
Research through Grant 2-1234-91-01.
\vskip0.5in
{\bf References}
\item{1}
L.D.Landau, E.M.Lifshitz, {\sl The Classical Theory of Fields}, Pergamon Press,
Oxford (1975), $\S$112.
\item{2}
A.Krasinski, {\sl The Physics in the Inhomogeneous Universe}, Warsaw University
Press (1993).
\item{3}
I.Dymnikova, R.de Ritis, C.Rubano, P.Scudellaro, to be published.
\item{4}
E.B.Gliner, {\sl Sov. Phys. JETP} {\bf 22}, 378 (1966).
\item{5}
A.Guth, {\sl Phys.Rev.} {\bf D23}, 389 (1981).
\item{6}
A.D.Sakharov, {\sl  Sov. Phys.} JETP {\bf 22}, 241 (1966).
\item{7}
E.B.Gliner, {\sl Sov. Phys. Dokl.} {\bf 15}, 559 (1970).
\item{8}
E.B.Gliner, I.G.Dymnikova, {\sl Sov. Astr. Lett.} {\bf 1}, 93 (1975).
\item{9}
A.D.Linde, {\sl Particle Physics and Inflationary Cosmology}, Harvard Academic
Press. Geneva (1990).
\item{10}
M.S.Turner, {\sl Inflation After COBE}, FERMILAB-Conf.92/313-A (1992).
\item{11}
L.H.Ford, {\sl Phys. Rev.} {\bf D31}, 710 (1985).
\item{12}
E.Mottola, {\sl Phys. Rev.} {\bf D33}, 1616 (1986).
\item{13}
C.Quigg, {\sl Gauge Theories of the Strong, Weak, and Electromagnetic
Interactions}, Addison-Wesley Publ. Co (1983); T.-P. Cheng and L.-F. Li,
{\sl Gauge Theory of Elementary Particle Physics}, Clarendon Press, Oxford
(1984).
\item{14}
P.D.B.Collins, A.D.Martin, E.J.Squires.., {\sl Particle Physics and Cosmology},
John Wiley and Sons, 1991
\item{15}
I.E.Dymnikova, {\sl Sov. Phys. JETP} {\bf 63}, 1111 (1986).
\item{16}
L.D.Landau, E.M.Lifshitz, {\sl Statistical Physics, Pergamon Press},
Oxford (1975).
\item{17}
V.Amaldi et al., {\sl Phys. Lett.} {\bf B260}, 447 (1991).
\vfill\eject

\end